**Treats or Affection? Understanding Reward Preferences in Indian Free-ranging Dogs for Bonding with Humans**

Srijaya Nandi[a], Aesha Lahiri[a], Tuhin Subhra Pal[a], Anamitra Roy[a], Rittika Bairagya[b], Anindita Bhadra[a*]

**Affiliations**

[a]Department of Biological Sciences, Indian Institute of Science Education & Research (IISER) Kolkata, Mohanpur, Nadia, West Bengal, India, PIN: 741246

[b]Department of Biotechnology, Maulana Abul Kalam Azad University of Technology, Haringhata, Nadia, West Bengal, India, PIN: 741249

*Corresponding Author

Email address: abhadra@iiserkol.ac.in (A. Bhadra).

**ORCID ID:**

Srijaya Nandi: https://orcid.org/0000-0002-9462-2374

Aesha Lahiri: https://orcid.org/0000-0003-0605-3827

Tuhin Subhra Pal: https://orcid.org/0000-0003-0829-9951

Anamitra Roy: https://orcid.org/0000-0002-3406-1667

Rittika Bairagya: https://orcid.org/0009-0007-0910-5932

Anindita Bhadra: https://orcid.org/0000-0002-3717-9732

**Abstract**

Free-ranging dogs constitute approximately 80% of the global dog population. These dogs are freely breeding and live without direct human supervision, making them an ideal model system for studying dog-human relationships. Living in close proximity with humans, free-ranging dogs in India frequently interact with people, and previous studies suggest that humans are a crucial part of their social environment. Positive reinforcement in the form of food and petting is commonly received from humans. In this study, we investigated which type of reward—food or petting—is more influential in shaping positive social associations with unfamiliar humans. Field trials were conducted on 61 adult free-ranging dogs in Nadia district, West Bengal, India. During a familiarization phase, two unfamiliar individuals each provided either food or petting to the dogs. This was followed by a choice test in which dogs could choose between the two individuals. A choice test following the familiarization phase was conducted on the first 5 days, while only choice tests were conducted on the subsequent 5 days. Results showed that on the first day, dogs significantly preferred the individual providing food. However, from the second day onward, preference for either individual was no different from chance. These findings suggest that while food is a stronger short-term motivator, repeated interactions involving either food or petting contribute equally to the formation of positive social associations over time. This study sheds light on the dynamics of dog-human relationship development in Indian free-ranging dog populations and highlights the nuanced role of different types of rewards in fostering affiliative bonds.

**Keywords**

*Canis lupus familiaris*, learning, dog-human relationship, food versus petting

**Introduction**

The domestic dog, *Canis lupus familiaris* (Linnaeus 1758), is the first species to have been domesticated by humans (Clutton-Brock, 1995) and is widely regarded as man's best friend (Ostrander and Giniger, 1997). It is believed that the domestication of dogs took place between 40,000 and 15,000 years ago, during the Last Glacial Maximum (LGM), in various regions of Europe and/or Asia (Larson et al., 2012; Ovodov et al., 2012; Frantz et al., 2016; Botigué et al., 2017; Bergstrom et al., 2020; Perri et al., 2021). Hunter-gatherers of the Pleistocene epoch are thought to have played a significant role in this process. While the domestication of approximately twenty animal species to date has primarily been for purposes such as food, clothing, and transportation (Beaver, 2009), the reasons and mechanisms behind the domestication of dogs remain unclear.

Since their domestication, dogs have occupied a special place in human society (Nagasawa et al., 2009), serving various purposes such as hunting, guarding, and herding in early human societies. In more recent times, they have taken on additional important roles, including service dogs, sniffer dogs, and companion animals (Coppinger and Schneider, 1995). Dogs have even become integral members of human households, forming a unique and special bond with our species. Several traits have contributed to their esteemed status, particularly their ability to respond to human gestures and language, as well as their hypersocial behaviour (Bhattacharjee et al., 2017). Caring for dogs evokes positive emotions in humans, that are similar to the bond shared between mothers and infants (Serpell, 2003). Likewise, dogs form attachments to their owners in much the same way children form attachments to their parents (Horn et al., 2013).

There is considerable uncertainty regarding the relative roles of different reinforcers in the development of social bonds between dogs and humans. Food is often cited as the most powerful reinforcer (Fukuzawa and Hayashi, 2013; Feuerbacher and Wynne, 2012; Elliot and King, 1960). However, some studies suggest that social interaction, such as petting, is just as effective, or even more so, in establishing bonds (Brodbeck, 1954; Igel and Calvin, 1960; Fonberg et al., 1981; Bhattacharjee et al., 2017). A study by Lazzaroni et al. (2020) found free-ranging dogs to show no clear preference between a person providing social contact and one offering food, suggesting that these dogs may not prioritize one form of reward over the other.

In the Global North, it is common to keep dogs as companion animals. However, the situation is quite different in countries of the Global South, such as India, where dogs are not only owned as pets, but are also found to live freely, sharing human habitats (Lord et al., 2013). These dogs, known as free-ranging, make up about 80% of the world's dog population. They live on the streets, breed freely, and are not under direct human supervision (Boitani and Ciucci, 1995; Bonanni and Cafazzo, 2014; Hughes and Macdonald, 2013; Serpell, 1995). They are primarily scavengers depended on human-generated waste, but they also beg and rely on handouts from humans for sustenance (Bhadra et al., 2016). Unlike pet dogs, which are products of artificial selection, free-ranging dogs are closer to their wolf-like ancestors (Akey et al., 2010; Shannon et al., 2015). In India, they have lived alongside humans for centuries (Thapar, 1990) and can be found in almost all human settlements, from rural areas to urban centers (Vanak and Gompper, 2009a). However, they often cause disruptions by scattering garbage, defecating in public spaces, barking at night, and acting as reservoirs for zoonotic diseases like rabies (Fekadu, 1982), making them a nuisance to local communities.

In the Indian society, there is a stark contrast in how people perceive and interact with dogs. While some people care for and regularly feed them, others subject them to violence, with incidents of dogs being beaten or killed by humans not being uncommon. In fact, humans are the leading cause of mortality for dogs, particularly in their early life (Paul et al., 2016). Given this, it is important for dogs to assess the intentions of unfamiliar humans before interacting with them.

Two common rewards that free-ranging dogs in India receive from humans are food and petting, with food being the more frequent reward (findings from ongoing surveys). In an earlier study, dogs provided with petting as an additional reward over several days showed more affiliation towards the experimenter as compared to dogs treated with additional food rewards. While this study revealed a population-level positive response to petting over food, the strength of this preference needs to be tested by providing both options to individual dogs. In this study, we aim to explore the differing roles that food and petting play in fostering social bonds between humans and free-ranging dogs, using a choice test, where each dog has the option to choose between two humans providing two kinds of rewards. Since free-ranging dogs are scavengers and frequently beg for food from humans, we hypothesized that food would be a stronger driver in the development of positive social bonds with humans.

**Methods**

**Subjects**

A total of 99 randomly selected adult free-ranging dogs were tested. To ensure that the dogs were adults, two criteria were applied: a male dog was considered an adult if his testes had descended, while a female was considered an adult if her nipples were dark in color. Residents of the areas where the dogs lived also confirmed that the dogs were adults (age ≥ one year). The dogs' sex was determined by examining their genitalia. Only visibly healthy dogs, with no signs of injury or illness, were selected for the experiments. Since dogs are typically social animals, most of the selected dogs were part of a group. Non-focal dogs in the group were lured away by the experimenter performing the video recording. All the dogs chosen for the experiment were unfamiliar to the experimenters. The dogs were photographed for tracking purposes, and their locations were marked using GPS on cell phones. They were identified based on their location, sex, body color, distinctive patches, ear and tail shape, and other unique features, such as scars from previous injuries.

**Study area**

The experiments were conducted in Gayeshpur (22°57'19.40"N, 88°29'45.88"E), Kataganj (22°57'2.61"N, 88°28'31.82"E) and Kalyani (22°58′30″N, 88°26′04″E) in Nadia district, West Bengal, India (Figure S1).

**Experimenters involved**

Three adults, all unfamiliar to the dogs, participated in the experiments. Two were females (Experimenter 1 [E1] and Experimenter 2 [E2]), and one was male (Experimenter 3 [E3]). E1 and E2 were responsible for giving rewards to the dogs and also participated in the choice tests by acting either as the food provider (FP) or petting provider (PP), while E3 performed the approachability test on the first day of the experiment (Day 0) and recorded videos of the experiments on the subsequent days.

**Experimental procedure**

**Approachability test**

The approachability test was conducted to assess the sociability of the dogs, which can be defined as their tendency to approach an unfamiliar experimenter (Bhattacharjee et al., 2021). On the first day of the experiment (Day 0), the experimenters walked through an area, and upon spotting one or more dogs, they conducted the test on individual dogs rather than groups. E3 stood approximately 4-5 meters from the dog and called out with a positive vocalization (ae-ae-ae) for up to 60 seconds. Since the dogs were free-ranging and not on a leash, E3 had to adjust his position in relation to the dog by eye estimation. E3 also maintained eye contact with the dog throughout the experiment. The dog was given a maximum of 60 seconds to approach E3 within about one body length (approximately 0.8m). Only dogs that approached E3 within approximately 0.8m and showed no signs of fear (e.g., crouched body, tail droop) or aggression (e.g., growling, barking, snarling) within 60 seconds on Day 0 were selected for further experimentation.

**Familiarization Phase**

The familiarization phase took place from Day 1 to Day 5. The primary goal of this phase was to help the dogs associate the experimenters (E1 and E2) with the specific reward they received from them. For each dog, one experimenter provided food rewards, while the other provided petting rewards across all five days. The assignment of rewards was randomized for each dog: for some, E1 provided food while E2 gave petting, and for others, the rewards were reversed. On any given day, both E1 and E2 gave their predetermined rewards to the dog, ensuring there was approximately 2-minute gap between each reward. The sequence in which the rewards were given was randomized across days.

**Familiarization involving Food**

The experimenter providing food, the food provider (FP) stood in front of the dog and called it using the same protocol as the approachability test. If the dog approached within 1 body length from the experimenter within 60 seconds, a piece of raw chicken (weighing approximately 8-10g) was dropped in front of the dog. The experimenter stood in front of the dog and gazed at it until the dog finished eating the chicken, after which the experimenter left and hid in an e-vehicle parked away from the dog.

**Familiarization involving Petting**

The protocol followed by the petting provider (PP) was the same as that followed by FP. If the dog approached within 1 body length from the experimenter within 60 seconds, the experimenter petted the dog by running the fingers of her right hand from the top of the dog's head to the neck and also gazed at it for approximately 10 seconds.

If the dog did not approach within approximately 0.8m of the experimenter within 60 seconds during either the food or petting trials, the trial was considered unsuccessful. A maximum of three food and three petting trials were allowed per dog per day. E3 remained near the dog the entire time.

**Test phase**

The test phase was conducted to assess the dogs' preference for the person providing food or petting rewards and involved a simple choice test involving E1 and E2. In each trial, E1 and E2 stood in front of the dog at a distance of approximately 4-5 meters, maintaining a distance of about 1-1.5 meters between each other. The dog, E1, and E2 were positioned in a triangular arrangement. The trial began with both experimenters calling the dog using the same vocalization which was used in the approachability test and the familiarization phase. They called for a maximum of 60 seconds, maintaining eye contact with the dog throughout this period. If the dog did not approach within one body length (approximately 0.8m) of the experimenters within 60 seconds, the trial was considered unsuccessful.

The test phase took place from Day 1 to Day 10, with a maximum of 3 trials per day. More than one trial per day was conducted only if the dog did not approach either of the experimenters or if the choice was unclear. From Days 1 to 5, the test phase followed the familiarization phase involving the food provider (FP) and the petting provider (PP). On Days 6 to 10, only the test phase was conducted. The time gap between the familiarization and test phase was 2 minutes.

To minimize potential bias toward a particular experimenter (FP or PP), they wore the same-coloured clothing and shoes throughout all phases of the experiment. The side on which FP and PP stood during the trials was also randomized across days. All experiments were conducted either in the morning (06:00-09:00 hours) or evening (15:00-18:00 hours) to align with the convenience of the experimenters, the activity patterns of the dogs, and the availability of daylight. Only daylight hours were selected for the experiments to ensure adequate visibility. The experimental protocol is outlined in Figure 1.

**Figure 1.** Schematic representation of the experimental protocol. Abbreviations: a) E1, E2, E3: Experimenter 1, 2 and 3, b) bl: body length of dog, i.e. the length from the nose tip to the tail base (1bl = ~0.8 m). Illustration by Arpan Bhattacharyya.

**Human flux estimation**

The human flux in the areas where the dogs were located was measured using the protocol outlined in Bhattacharjee et al. (2021). Flux videos were recorded by E3 only for those locations where the dogs completed the approachability test conducted on Day 0. A total of 12 flux videos were recorded for each location where the dogs lived, across four randomly selected days. Two videos were recorded during the 06:00-09:00 hours slot, and the remaining two were recorded during the 15:00-18:00 hours slot. The mean flux for each location was calculated using these 12 videos. Due to an imbalance in the number of dogs tested in the low and intermediate flux areas (more dogs belonged to low flux areas), we avoided classifying areas into the different flux categories. Instead, the absolute values of flux obtained after averaging were used for analysis. High flux areas are characterized by heavy vehicle traffic, which poses a significant risk to free-ranging dogs, often resulting in fatal accidents. To minimize the likelihood of losing subjects during the experiment, we deliberately excluded dogs from high flux areas.

**Video decoding and analysis**

All the videos were coded by S.N., and the obtained data were used for further analysis. Only the dogs that completed and showed a clear choice in all ten choice tests (up to Day 10) were used for analysis. We checked four parameters across different days of the experiment which were namely choice, latency of approach (in seconds), trial number in which the dog approached and socialization index (SI).

**Choice:** A choice was scored if a focal dog approached close to i.e., within one body length (approximately 0.8m) of the experimenter and at least gazed at her during the choice tests.

**Trial number:** This variable represented the sequential number of the trial in which the dog approached either the familiarization phase or the test phase (choice test). Trial number could have a maximum of three levels (1, 2 and 3 representing the first, second and third trial, respectively). This parameter was however, not considered in the analysis due to the very infrequent occurrence of trial numbers 2 and 3, both in familiarization and test phases.

**Approach latency:** Approach latency was measured in seconds and was defined as the time taken by the dogs to approach within one body length of the experimenter after the dog noticed the experimenter for the first time.

**Socialization Index (SI):** The Socialization Index (SI) was scored following the method described in Table 1 of Nandi et al. (2023), with the addition of one new behaviour observed in the current study (Table 1).

**Table 1.** Behaviours exhibited by the tested dogs and their corresponding scores used to calculate the Socialization Index (SI).

| Behaviour | Score |
|---|---|
| No tail wag | 0 |
| Slow tail wag | 1 |
| Fast tail wag (without back movement) | 2 |
| Rapid tail wag (with back movement) | 3 |
| Licking experimenter | 4 |
| Pawing experimenter (using either the left or right paw to touch the experimenter) / Nudging experimenter (using nose to touch the experimenter) | 5 |
| Jumping on experimenter's body / Affiliative vocalizations while looking at the experimenter | 6 |

Choice of the food provider and petting provider was coded as 1 and 0, respectively. A generalized linear model (GLM) with a binomial distribution was used to model choice behaviour.

Approach latency was analyzed using a Cox proportional hazards model when the proportional hazards assumption was met. In cases where this assumption was violated, an accelerated failure time model was used.

The SI ranged from 0 to 32. To normalize the scores, each SI value was divided by the maximum possible SI score (32), scaling it to a range between 0 and 1. Since the number of zero values was minimal, a small non-zero value was added to ensure all transformed SI values fell strictly within this range, allowing for a beta regression. Effect size and associated confidence intervals was mentioned for different parameters. Model-predicted estimated marginal means was also reported along with the confidence intervals.

All statistical analyses were conducted using R Studio (R Development Core Team, 2022). An alpha level of 0.05 was used for all the analyses conducted.

**Results**

Of the 99 dogs that participated in the approachability test on Day 0, 61 (30 females, 31 males) approached within one body length of E3 and were studied over the following 10 days. Among them, 51 dogs (27 females, 24 males) completed all trials, while 45 (21 females, 24 males) not only completed all trials but also demonstrated a clear preference between the food provider (FP) and petting provider (PP). Only these 45 dogs, who consistently made a clear choice across all 10 trials, were included in the final analysis.

**A. Approachability test (Day 0)**

**A1. Effect of dog sex, flux and SI on approach latency**

On Day 0, the overall approach latency was 11.689 ± 12.901 seconds. Males approached in 11.083 ± 13.590 seconds on average, while females took 12.381 ± 12.363 seconds. A Cox proportional hazards model was fitted to examine the effects of dog sex, flux, and SI on approach latency. The analysis included 45 observations, all of which recorded an event (i.e., all dogs approached; no censored cases). Assumptions of proportional hazards were tested and met (all $p > 0.05$; Table S1). None of the predictors significantly influenced approach latency (all $p > 0.05$; Table S2). The model fit was not significant (Likelihood ratio test: $\chi^2(3) = 1.10$, $p = 0.80$), and the concordance index was 0.573, indicating limited predictive ability. The model was specified as:

$$\text{model} = \text{approach\_latency} \sim \text{sex} + \text{flux} + \text{si\_day0}$$

**A2. Effect of dog sex, flux and approach latency on SI**

The overall SI on Day 0 was 1.844 ± 1.678. The values for males were 1.333 ± 0.761 and for females were 2.428 ± 2.204.

A beta regression model was fitted to assess the effect of sex, flux and approach latency on SI. Sex was found to significantly influence SI ($\beta$ = -0.600, SE = 0.253, p = 0.018, exp($\beta$) = 0.549, 95% CI: 0.334 -0.901), with females showing approximately 1.8 times the SI as that of males. Consequently, the model-estimated marginal mean SI for females on the response scale was 0.073 (95% CI: 0.052 – 0.102), while for males, it was 0.042 (95% CI: 0.028 – 0.062), reflecting a higher estimated SI for females. However, approach latency and flux were not statistically significant in predicting SI (Table S3). The pseudo $R^2$ was 0.089, indicating that fixed effects explained 8.9% of the variance in the response variable. The model's prediction error, as measured by RMSE was 0.051, indicating a reasonable fit. The model specified was:

$$model = new\_si \sim approach\_latency + sex + flux$$

**B. Familiarization phase followed by test phase (Day 1)**

**B1. Do FRDs prefer the food or petting provider on Day 1?**

A chi-square goodness-of-fit test was performed to check whether there was any preference among the 45 dogs tested between FP and PP on Day 1. Results indicated a significant preference for FP compared to PP on Day 1 ($\chi^2$ = 5.00, *p* = 0.025). Choice on Day 1 is given in Figure 2a.

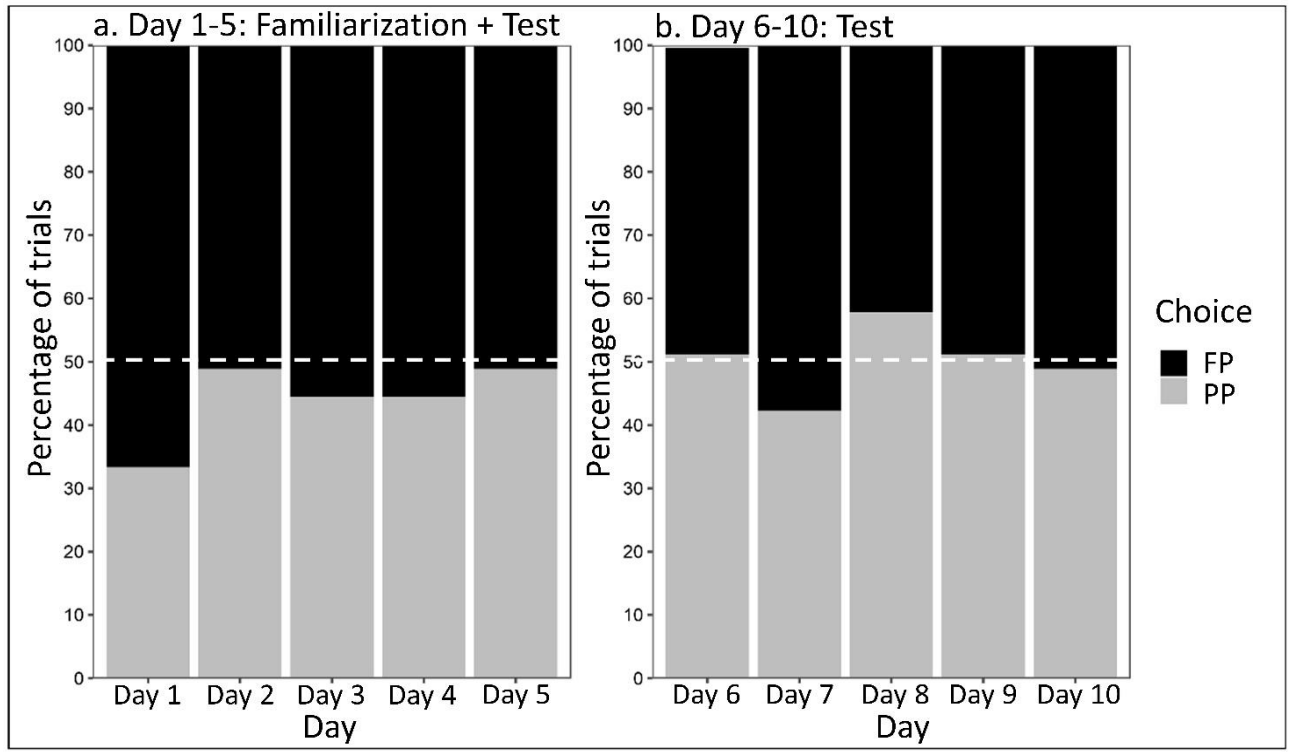

**Figure 2.** A bar graph representing percentage of trials in which food provider (FP) and petting provider (PP) were chosen during the test phase conducted from **a)** Day 1 to 5: when the test phase was conducted following the familiarization phase and **b)** Day 6 to 10: when only the test phase was conducted.

**B2. What factors affect the person chosen on Day 1?**

A generalized linear model with a binomial distribution was performed with the person chosen as the outcome variable and sex, flux, SI towards PP and FP (during the familiarization phase), sequence in which the reward was given and side chosen as predictor variables. None of the predictors was significant in determining the person chosen (Table S4). The model specified was:

$$\text{model} = \text{choice} \sim \text{sex} + \text{flux} + \text{si\_pp} + \text{si\_fp} + \text{sequence} + \text{side}$$

**C. Familiarization phase followed by test phase (Day 1 to 5)**

**C1) What factors affect approach latency during the familiarization phase from Day 1 to 5?**

First, a Cox proportional hazards model was run with sex, flux, day, person (PP or FP) involved, SI displayed, sequence of visit of reward provider, and approach latency on Day 0 as predictor variables, with approach latency during the familiarization phase as the response variable. Dogs that did not approach either the PP or FP during Trial 1 of the familiarization phase were censored. This model violated the proportional hazards assumption (Table S5).

A log-logistic accelerated failure time (AFT) model was then fitted which directly models survival time rather than hazard rates and does not assume proportional hazards. The model revealed a significant negative association between the SI ($\beta = -0.043$, SE = 0.012, p <0.001, TR = 0.958, 95% CI: 0.935-0.981) and approach latency, indicating that higher SI values were associated with faster approach times. Similarly, approach latency significantly decreased over time, with Day 2, Day 3, Day 4 and Day 5 (Table 2) showing progressively shorter latencies compared to Day 1. Other predictors, including sex, reward provider, flux, sequence, and approach latency on Day 0, were not significant in determining approach latency during familiarization phase from Day 1 to 5 (Table S6). The estimated scale parameter was 0.336, and the model fit was significant ($\chi^2=107.39$, df = 10, p<0.001). These results suggest that socialization and repeated exposure over days strongly influenced approach latency, while other factors had no impact. A Kaplan-Meier survival curve illustrating approach time during familiarization phase from Day 1 to 5 is presented in Figure 3a. The model specified was:

model = approach_latency ~ sex + flux + person + si + approach_latency_day0 + day + sequence

**Table 2.** Results of an accelerated failure time (AFT) model predicting survival time during the familiarization phase conducted from Day 1 to 5.

| Term  | Estimate | Std. Error | p      | TR    | 95% CI      |
|-------|----------|------------|--------|-------|-------------|
| Day 2 | -0.307   | 0.091      | <0.001 | 0.735 | 0.615-0.879 |
| Day 3 | -0.590   | 0.087      | <0.001 | 0.554 | 0.467-0.658 |
| Day 4 | -0.620   | 0.087      | <0.001 | 0.538 | 0.453-0.638 |
| Day 5 | -0.760   | 0.087      | <0.001 | 0.468 | 0.394-0.554 |

**C2) What factors affect SI during the familiarization phase from Day 1 to 5?**

A generalized linear model with beta distribution was fitted with SI displayed as the response variable and sex, flux, day, person (PP or FP) involved, sequence of visit of reward provider, and SI on Day 0 as predictor variables.

The beta regression model predicting the SI showed a significant effect of several predictors. Compared to Day 1, dogs displayed approximately 1.5 times more SI on Day 2, 1.5 times more SI on Day 3, 1.7 times more SI on Day 4, and 1.4 times more SI on Day 5 (Table 3). Pairwise comparisons revealed no significant differences among Days 2, 3, 4, and 5 (all p-values > 0.05 after adjustment), suggesting that SI remained relatively stable after the initial increase from Day 1. SI on Day 0 was positively associated with the displayed SI during the familiarization phase ($\beta = 0.091$, $SE = 0.020$, $p < 0.001$, $\exp(\beta) = 1.095$, 95% CI: 1.053 – 1.139), indicating that one-unit increase in SI on Day 0 made dogs approximately 1.1 times more likely to display a higher SI during the familiarization phase. Sequence in which the reward provider visited the dog had a significant influence on SI ($\beta = -0.159$, $SE = 0.074$, $p = 0.033$, $\exp(\beta) = 0.853$, 95% CI: 0.737 – 0.987). SI displayed towards the person visiting the second was 0.8 times as that for the first. Flux, sex and person did not have significant effects (Table S7). The pseudo $R^2$ was 0.087, indicating that fixed effects explained 8.7% of the variance in the response variable. Model prediction error was 0.071 RMSE, indicating a reasonable fit. The effects of SI on Day 0, sequence in which the reward was given and the day of experiment is presented in Figure 3b, 3c and 3d respectively.

Consequently, the model-estimated marginal mean new_SI for Day 1 on the response scale was 0.057 (95% CI: 0.048 – 0.068), for Day 2 was 0.081 (95% CI: 0.069 – 0.094), for Day 3 was 0.085 (95% CI: 0.073 – 0.099), for Day 4 was 0.092 (95% CI: 0.079 – 0.106) and for Day 5 was 0.079 (95% CI: 0.067 – 0.092). SI displayed for the person visiting first was 0.084 (95% CI: 0.076 – 0.093), while for the one visiting later was 0.072 (95% CI: 0.065 – 0.081). The model specified was:

model = si_new ~ sex + person + sequence + flux + si_day0 + day

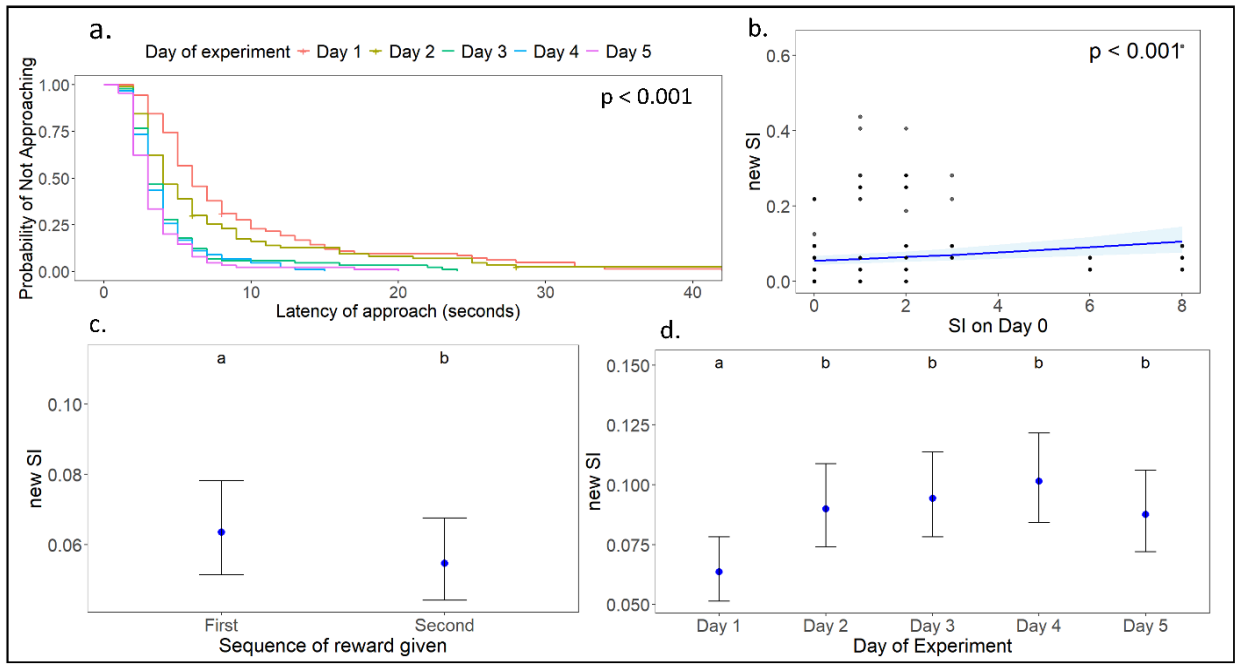

**Figure 3.** Visual summary of the familiarization phase conducted from Day 1 to Day 5: **(a)** Kaplan–Meier survival curves illustrating the time to approach (seconds). Statistical comparison across days was performed using the log-rank test ($p < 0.001$), **(b)** Scatterplot showing the effect of SI displayed on Day 0 on new SI (SI normalized to lie between 0 and 1). The blue line represents the linear regression line, and the blue area depicts its 95% CI, illustrating the uncertainty around the estimated relationship, **(c)** Effect plot showing the impact of sequence in which reward was given on new SI, **(d)** Effect plot showing the impact of day of experiment on new SI. The dots represent the model-fitted mean while the whiskers represent the uncertainty (95% CI).

**Table 3.** Results of a beta regression model predicting SI displayed during familiarization phase from day 1 to 5.

| Term | Estimate | Std. Error | p | exp(estimate) | 95% CI |
| --- | --- | --- | --- | --- | --- |
| Day 2 | 0.376 | 0.122 | 0.002 | 1.456 | 1.146-1.850 |
| Day 3 | 0.430 | 0.122 | <0.001 | 1.537 | 1.210-1.952 |
| Day 4 | 0.509 | 0.121 | <0.001 | 1.664 | 1.313-2.110 |
| Day 5 | 0.347 | 0.123 | 0.005 | 1.415 | 1.111-1.801 |

**C3. What factors affect approach latency during the test phase from Day 1 to 5?**

A Cox proportional hazards model was fitted to examine the effect of sex, flux, day, SI, approach latency on Day 0 and reward chosen on the approach latency during test phase from Day 1 to 5. The model included 225 observations and 220 events. The test for proportional hazards assumption indicated some possibility of violation (Table S8).

A log-logistic accelerated failure time (AFT) model was then fitted which directly models survival time rather than hazard rates and does not assume proportional hazards. The model revealed a significant negative association between the SI ($\beta$ = -0.070, SE = 0.019, p <0.001, TR = 0.933, 95% CI: 0.897, 0.969) and approach latency, indicating that higher SI values were associated with faster approach times. Similarly, approach latency significantly decreased over time, with Day 2, Day 3, Day 4 and Day 5 showing progressively shorter latencies compared to Day 1 (Table 4). Other predictors, including sex, person chosen, flux and approach latency on Day 0, were not significant in determining approach latency. The estimated scale parameter was 0.372, and the model fit was significant ($\chi^2$=44.02, df = 9, p<0.001). These results suggest that socialization and repeated exposure over days strongly influenced approach latency, while other factors had no impact (Table S9). A Kaplan-Meier survival curve illustrating approach time during the test phase from Day 1 to 5 is presented in Figure 4a. The model specified was:

$$\text{model} = \text{approach\_latency} \sim \text{sex} + \text{flux} + \text{choice} + \text{approach\_latency\_day0} + \text{day} + \text{si}$$

**Table 4.** Results of an accelerated failure time (AFT) model predicting approach latency during test phase from Day 1 to 5.

| Term | Estimate | Std. Error | p | TR | 95% CI |
| --- | --- | --- | --- | --- | --- |
| Day 2 | -0.429 | 0.142 | 0.002 | 0.651 | 0.493-0.860 |
| Day 3 | -0.478 | 0.138 | <0.001 | 0.620 | 0.473-0.813 |
| Day 4 | -0.663 | 0.140 | <0.001 | 0.515 | 0.391-0.679 |
| Day 5 | -0.693 | 0.140 | 0.005 | 0.500 | 0.380-0.658 |

## C4. What factors affect SI during the test phase from Day 1 to 5?

A beta regression model was fitted to predict si_new using sex, SI displayed on Day 0, flux, day, and person chosen as predictors. Among the predictors, SI on Day 0 had a significant positive effect (β = 0.094, SE = 0.026, $p$ <0.001), suggesting that an increase in SI on Day 0 was associated with higher si_new. Flux, sex, reward chosen and the day of the experiment did not have significant effects on si_new (Table S10). The effect of SI on Day 0 is presented in Figure 4b.

The model specified was:

$$model = si\_new \sim sex + flux + choice + day + si\_day0$$

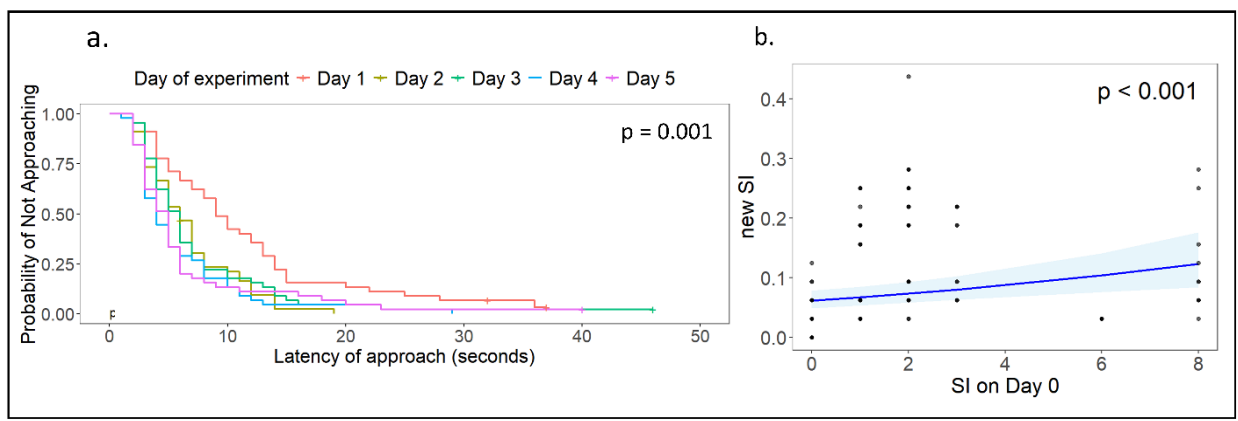

**Figure 4.** Visual summary of the test phase conducted from Day 6 to Day 10: **(a)** Kaplan–Meier survival curves illustrating the time to approach (seconds). Statistical comparison across days was performed using the log-rank test ($p < 0.001$). **(b)** Scatterplot showing the effect of SI displayed on Day 0 on new SI (SI normalized to lie between 0 and 1). The blue line represents the linear regression line, and the blue area depicts its 95% CI, illustrating the uncertainty around the estimated relationship.

## C5. Are dogs consistent in their choice from Day 1 to 5?

Fleiss' Kappa was used to assess the consistency of choices across trials. The analysis was conducted across 5 days and there were two possible choices (FP and PP). The agreement was found to be κ = -0.22, indicating lower agreement than expected by chance. A z-score of -1.06 was obtained, with a p-value of 0.29, suggesting that the observed agreement was not statistically significant. This result implies a lack of consistency in choices across the test phase conducted from Day 1 to 5. Figure 2a shows the choice displayed by the dogs from Day 1 to 5.

**C6. What factors affect choice from Day 1 to 5?**

A generalized linear model (GLM) was fitted using a binomial distribution with a logit link function to examine the effect of various predictors on the likelihood of choosing the reward provider. The model included sex, flux, day, SI towards PP, SI towards FP, sequence of visit of reward provider and side chosen as fixed effects.

Results indicated higher SI displayed towards PP to significantly decrease the probability of choosing FP ($\beta$ = -0.151, SE = 0.061, p = 0.013, OR = 0.859, 95% CI: 0.763-0.969). Also, a left side bias was observed ($\beta$ = -0.758, SE = 0.285, p = 0.008, OR = 0.469, 95% CI: 0.268-0.820). The other predictors did not show significant effects (Table S11). The model specified was:

$$\text{model} = \text{choice} \sim \text{sex} + \text{flux} + \text{day} + \text{side} + \text{sequence} + \text{si\_pp} + \text{si\_fp}$$

**D. Test phase (Day 6 to 10)**

**D1. What factors affect approach latency during the test phase from Day 6 to 10?**

The effect of sex, flux, approach latency on Day 0, the day of the experiment, reward provider chosen (FP or PP) and the SI displayed were checked for their impact on the approach latency displayed during the test phase from Day 6 to 10 using a Cox proportional hazards model. The test for proportional hazards assumption showed a violation (Table S12).

A log-logistic accelerated failure time (AFT) model was then fitted which directly models survival time rather than hazard rates and does not assume proportional hazards. The model revealed a significant negative association between the SI ($\beta$ = -0.064, SE = 0.019, p=0.001, TR = 0.938, 95% CI: 0.903-0.975) and approach latency, indicating that higher SI values were associated with faster approach times. Similarly, approach latency significantly increased over time, however, it was significantly higher only on Day 10 compared to Day 6 ($\beta$ = 0.469, SE = 0.148, p=0.002, TR = 1.598, 95% CI: 1.195-2.136). Approach latency on Day 0, sex, person chosen and flux were not significant in determining approach latency (Table S13). The estimated scale parameter was 0.402, and the model fit was significant ($\chi^2$=26.23, df = 9, p=0.002). A Kaplan-Meier survival curve illustrating approach time across Day 6 to 10 is presented in Figure 5a. The model specified was:

$$\text{model} = \text{approach\_latency} \sim \text{sex} + \text{choice} + \text{si} + \text{flux} + \text{approach\_latency\_day0} + \text{day}$$

**D2. What factors affect SI during the test phase from Day 6 to 10?**

The effect of dog sex, flux, reward provider chosen (FP or PP), day of the experiment (Day 6 to 10) and SI displayed on Day 0 were checked for their ability to predict the SI displayed by the dogs during the test phase from Day 6 to 10. Due to a near-zero variance across dogs, a standard beta model was fitted.

Sex had a significant effect, with males having lower SI scores than females ($\beta$ = -0.319, SE = 0.101, $p$ = 0.002, exp($\beta$) = 0.727, 95% CI: 0.596 – 0.886). SI on Day 0 was positively associated with SI displayed from Day 6 to 10 ($\beta$ = 0.106, SE = 0.025, $p$ < 0.001, exp($\beta$) = 1.112, 95% CI: 1.059 – 1.168), indicating consistency over time. Flux also had a significant negative effect ($\beta$ = -0.014, SE = 0.006, $p$ = 0.028, exp($\beta$) = 0.986, 95% CI: 0.974 – 0.998), indicating decreasing values of SI with an increase in flux. However, day and the reward person chosen did not have significant effects ($p$ > 0.05; Table S14). The effects of sex, flux, and SI on Day 0 are presented in Figure 5b, 5c, and 5d, respectively.

The estimated marginal means analysis on the response scale revealed that the SI displayed by females (EMM = 0.116, SE = 0.007, 95% CI: 0.102 - 0.130) was higher than males (EMM = 0.087, SE = 0.006, 95% CI: 0.076 – 0.099). The model specified was:

$$\text{model} = \text{si\_new} \sim \text{sex} + \text{flux} + \text{day} + \text{si\_day0} + \text{choice}$$

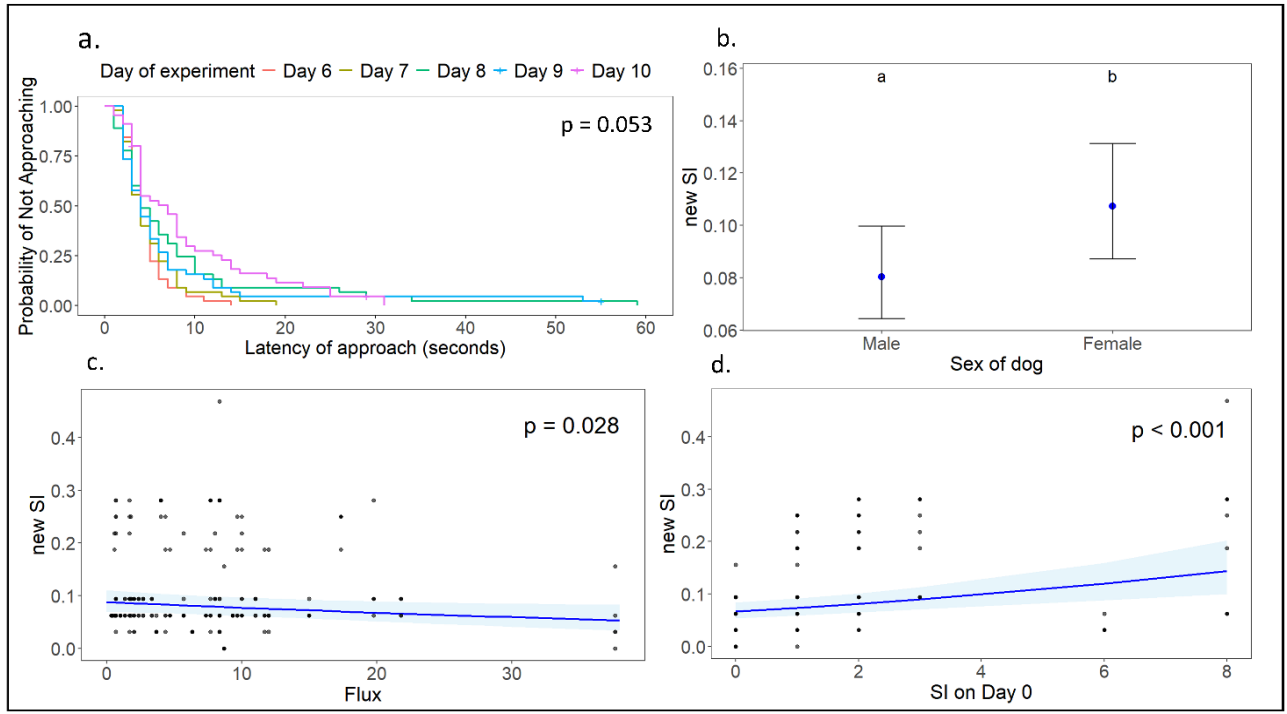

**Figure 5. (a)** Kaplan–Meier survival curves illustrating the time to approach (seconds). Statistical comparison across days was performed using the log-rank test ($p < 0.001$), **(b)** Effect plot showing the impact of sex of dog on new SI. The dots represent the model-fitted mean while the whiskers represent the uncertainty (95% CI), **(c)** Scatterplot showing the effect of flux on new SI (SI normalized to lie between 0 and 1), **(d)** Scatterplot showing the effect of SI displayed on Day 0 on new SI (SI normalized to lie between 0 and 1). The blue line represents the linear regression line, and the blue area depicts its 95% CI, illustrating the uncertainty around the estimated relationship.

**D3. Are dogs consistent in their choice from Day 6 to 10?**

Fleiss' Kappa was used to assess the consistency of choices. The analysis was conducted across these 5 days, and there were two possible choices (FP and PP). The agreement was found to be κ = -0.351, indicating lower agreement than expected by chance. A p-value of 0.18 suggests that the observed agreement was not statistically significant. This result implies a lack of consistency in choices across the test phase from Day 6 to 10. Figure 2b shows the choice displayed by the dogs from Day 6 to 10.

**D4. What factors affect choice from Day 6 to 10?**

The effect of sex, flux, day and side chosen were checked for their effect on person chosen from Day 6 to 10. A generalized linear model (GLM) was performed (due to very low variance across dogs). This model showed no significant effect of any of the predictors (Table S15). The model specified was:

$$model = choice \sim sex + flux + day + side$$

**Discussion**

Attachment in humans is often assessed through proximity-seeking behaviours (Bowlby, 1973). A similar concept applies to dogs, where proximity to a human can serve as an indicator of sociability (Barrera et al., 2010). The time taken to approach a human reflects the level of interest exhibited by dogs (Bhattacharjee et al., 2017), while their specific behaviours towards the experimenter may provide insights into the nature of their social bonds.

In the present study, approach latency to an unfamiliar person on Day 0 was not influenced by the dog's sex. However, SI values were affected, with male dogs exhibiting lower SI values than females. The Socialization Index (SI) is a measure of the sociability of the dogs towards humans. Hence, the results indicate that while the initial level of interest in approaching an unfamiliar human does not differ between the sexes, the propensity to engage in social interaction does. This finding aligns with previous studies that have reported sex-based differences in sociability with humans in pet dogs (Lore and Eisenberg, 1986; Scandurra et al., 2018).

Flux—a measure of human movement in the dog's environment—did not influence approach latency or SI toward the unfamiliar human. This contrasts with the findings of Bhattacharjee et al. (2021), who reported greater sociability in dogs from high and intermediate flux areas compared to those from low flux areas. The discrepancy may be attributed to differences in the experimental design: the current study included only dogs that approached the unfamiliar person on Day 0, thereby excluding such individuals who would be very low in sociability. The previous study assessed a mixed group that included both responsive and unresponsive individuals, and obtained a mean response of 34.5% in low flux zones, as opposed to 75.5% and 78.5% in high and intermediate flux zones respectively. Moreover, the lack of correlation between SI and approach latency on Day 0 suggests that a higher level of interest does not necessarily translate into greater sociability toward an unfamiliar human.

When reward preference was assessed, dogs showed a significant initial preference for the food provider (FP) on Day 1, when both individuals were unfamiliar. This contrasts with Lazzaroni et al. (2020), who reported no preference between food and cuddle providers on Day 1, potentially due to differences in geographic context or reward type. In the current study, raw chicken—a rarely provisioned but preferred food among free-ranging dogs (Bhadra et al., 2016) was used, possibly explaining the initial preference. From Day 2 onwards, as interactions with both providers increased, the dogs' choices became random. This suggests that while food served as a strong short-term motivator, repeated positive interactions with either provider eventually led to a lack of preference, thereby confirming the importance of petting as a reward for the free-ranging dogs.

No side bias was observed on Day 1, but it emerged on subsequent days and influenced choices until Day 5, before disappearing again from Day 6 to 10, when rewards were no longer provided. The cause of this disappearance remains unclear and warrants further investigation. Furthermore, choice between providers from Day 1 to 5 was significantly influenced by the SI displayed toward the petting provider during the familiarization phase, with higher SI values predicting a greater likelihood of choosing her during the choice tests. This suggests that more sociable dogs might have a higher, albeit less skewed, preference for petting as a reward.

Approach latency and SI during the familiarization phase (Days 1–5) showed significant temporal changes. Approach latency decreased while SI increased substantially from Day 1 to Day 2, after which both stabilized. These findings are in line with Nandi et al. (2024), who reported decreased approach latency after a single interaction and increased SI following two interactions. The faster rise in SI in the present study may be due to methodological differences. Notably, approach latency during familiarization was not predicted by latency toward the unfamiliar person on Day 0. However, SI during familiarization was positively correlated with SI toward the unfamiliar person on Day 0, suggesting that SI may be a more stable behavioural trait. The sequence of visits by the reward providers did not affect approach latency, but SI was lower for the second visitor, regardless of the reward type, possibly reflecting reduced enthusiasm during subsequent interactions.

Additionally, dogs with higher SI demonstrated shorter approach latencies, reinforcing the link between sociability and approach motivation. This, however, contradicts Day 0 findings, where SI and approach latency were not negatively associated. This is likely due to familiarity with the experimenters on Day 1, compared to Day 0 and/or the effects of positive reinforcement. Neither sex nor flux influenced approach latency or SI during this phase, diverging from Day 0 results where males were found to be less sociable, reinforcing the suggestion of familiarity and rewards influencing latency. Furthermore, no differences in approach latency or SI were observed between the two reward providers, indicating comparable motivation to engage with both.

During the test phase (Days 1–5), when choice tests followed the familiarization phases, approach latency continued to decrease from Day 1 to Day 2, following trends seen during familiarization. However, SI did not change significantly across these days, potentially due to a decline on Day 1 (immediately after the familiarization sessions), leading to a level of saturation.

From Days 6 to 10, when no rewards were offered and choices were based on memory, dogs maintained low approach latencies, except on Day 10, when latency increased significantly. This suggests a gradual reduction in motivation in the absence of rewards, a pattern consistent with reward extinction in other species (Skinner, 1938). Despite this, SI remained high throughout, indicating a persistent tendency to engage with familiar humans. This is consistent with anecdotal reports of Indian free-ranging dogs maintaining affiliative behaviours toward familiar individuals even after long periods without interaction or reinforcement.

Male dogs continued to show lower SI than females during Days 6–10, in line with earlier observations. Interestingly, flux emerged as a significant negative predictor of SI during this reward-absent phase—dogs from lower flux areas exhibited higher SI. Bhattacharjee et al. (2021) found that dogs from low-flux areas were more anxious around unfamiliar humans. In this study, the experimenters were no longer unfamiliar to the dogs in the reward-absent phase. Also, as mentioned earlier, in this study we included only those dogs that responded to the experimenters on Day 0, thereby considering a pool of less anxious individuals. Although flux did not affect SI on Day 0, its influence during the reward-absent phase suggests that dogs from areas with less human activity may have a greater drive for social contact, but may be more reluctant to approach unfamiliar humans. This would lead to more sustained affiliative behaviour in such dogs once familiarized.

Overall, dogs did not exhibit a consistent preference for either reward provider throughout the experiment. This inconsistency may be attributed to two factors: (1) both types of rewards became equally valued over time, and (2) dogs became habituated to the experimental setup and ceased to differentiate meaningfully between providers. Nonetheless, the strong initial preference for the FP underscores the powerful short-term motivational role of food.

Our findings contrast with those of Bhattacharjee et al. (2017), who reported that free-ranging dogs preferred petting over food in building trust during repeated interactions with unfamiliar humans. This discrepancy likely reflects differences in experimental paradigms. In the earlier study, food was the primary reward and could be accessed either from the experimenter's hand or the ground, with petting or food given as an additional motivator prior to the choice test. The decision of the dogs to eat from the hand was considered at trust. In contrast, this study used two distinct experimenters providing separate rewards, presenting a different context for reward-based preference. The choice of the dog was determined by its approach towards an experimenter within one body length and gazing towards the experimenter, which would require a higher level of motivation and sociability. There was no requirement to make actual contact with the experimenters in this paradigm. Further, the earlier study was conducted in separate sets of dogs, and the individual dogs did not have a choice between the two kinds of rewards. Hence, the earlier study reflects the population-level response to petting in free-ranging dogs, while this study tests the hierarchy of preference of individual dogs for reward givers of two types.

These findings offer valuable insights into the social behaviour of Indian free-ranging dogs, underscoring the nuanced interplay between immediate and long-term reinforcement in shaping social preferences. Future research should aim to explore additional ecological and individual-level factors that modulate reward preferences in dog–human interactions.

**Availability of data and materials**

Supplementary data associated with this article can be found in the online version at

**Acknowledgements**

The authors are thankful to Siddharth S. for his contribution in data collection and Mr. Rohan Sarkar, Dr. Udipta Chakraborty and Dr. Rubina Mondal for their valuable inputs in data analysis. We thank Mr. Arpan Bhattacharyya for creating illustrations of the experiment.

**Funding**

This research was supported by the Department of Science and Technology, Ministry of Science and Technology (INSPIRE fellowship to S.N.) and the Janaki Ammal – National Women Bioscientist Award (BT/HRD/NBA-NWB/39/2020-21 (YC-1)), Department of Biotechnology, India. PhD fellowship was granted to T.S.P. by UGC.

**Ethical considerations**

Only non-invasive procedures were used in the experiments. The protocol followed was very similar to the method used by people on the streets of West Bengal to call and interact with the free-ranging dogs and therefore elicited minimum disturbance to the animals studied.

**CRediT authorship contribution statement**

**Srijaya Nandi:** Conceptualization, Methodology, Formal analysis, Investigation, Data curation, Writing – Original Draft, Visualization, Project administration. **Aesha Lahiri:** Investigation. **Tuhin Subhra Pal:**

Investigation. **Anamitra Roy:** Investigation. **Rittika Bairagya:** Investigation. **Anindita Bhadra:** Conceptualization, Methodology, Resources, Writing – Review & Editing, Supervision, Funding acquisition.

**Declaration of Competing Interest**

The authors declare that they have no known competing financial interests or personal relationships that could have appeared to influence the work reported in this paper.

**Supplementary material**

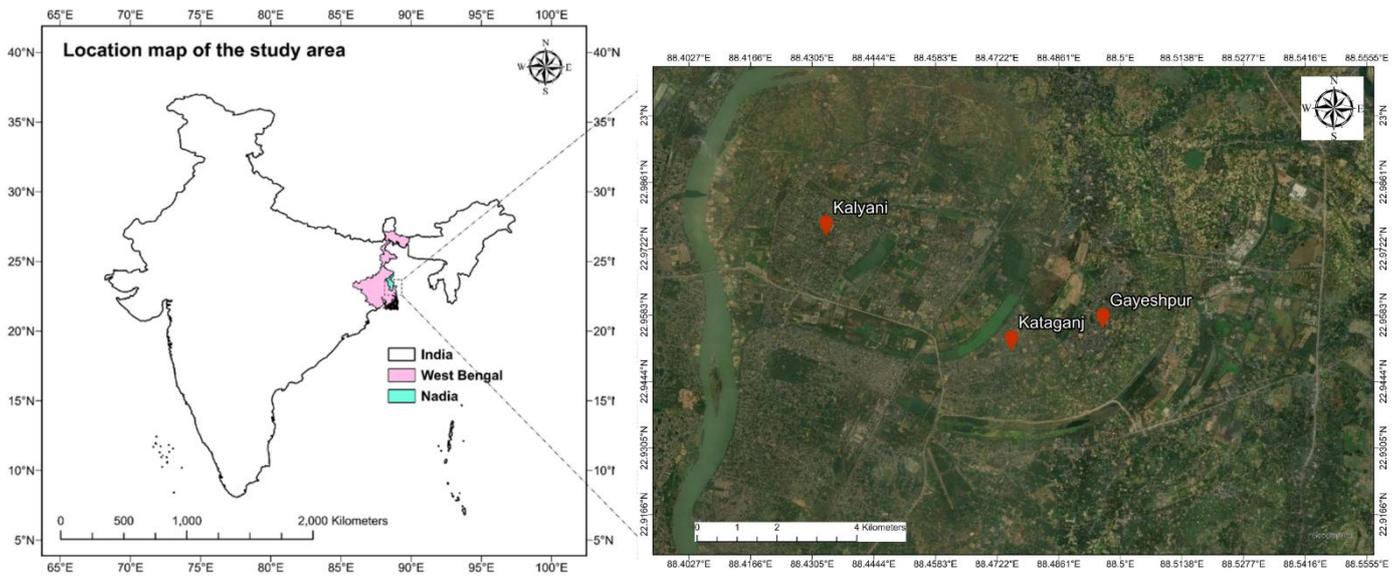

**Figure S1.** Study sites

**A1. Effect of dog sex, flux and SI on approach latency**

model = approach_latency ~ sex + flux + si_day0

**Table S1.** Results of test of proportional hazards assumption for Cox proportional hazards model performed for predicting approach latency on Day 0.

| Term | chisq | df | p |
| --- | --- | --- | --- |
| sex | 1.087 | 1 | 0.30 |
| flux | 0.955 | 1 | 0.33 |
| si_day0 | 0.474 | 1 | 0.49 |
| GLOBAL | 2.091 | 3 | 0.55 |

**Table S2.** Results of a Cox proportional hazards model predicting approach latency on Day 0.

| Term | coef | exp(coef) | se(coef) | z | Pr(>|z|) | exp(-coef) | lower .95 | upper .95 |
| --- | --- | --- | --- | --- | --- | --- | --- | --- |
| sexM | 0.006 | 1.006 | 0.334 | 0.018 | 0.986 | 0.994 | 0.523 | 1.934 |
| flux | 0.018 | 1.018 | 0.019 | 0.950 | 0.342 | 0.982 | 0.981 | 1.057 |
| si_day0 | -0.044 | 0.957 | 0.086 | -0.513 | 0.608 | 1.045 | 0.809 | 1.132 |

## A2. Effect of dog sex, flux and approach latency on SI

**Table S3.** Results of a beta regression model predicting SI on Day 0.

| Term | Estimate | Std. Error | z value | Pr(>|z|) |
|---|---|---|---|---|
| (Intercept) | -2.300 | 0.244 | -9.420 | <2e-16 |
| sexM | -0.600 | 0.253 | -2.370 | 0.018 |
| approach_latency_day0 | -0.003 | 0.010 | -0.355 | 0.722 |
| flux | -0.028 | 0.018 | -1.562 | 0.118 |

## B2. What factors affect the person chosen on Day 1?

model = choice ~ sex + flux + si_pp + si_fp + sequence + side

**Table S4.** Results of a generalized linear model with binomial distribution predicting choice of reward on Day 1.

| Term | Estimate | Std. Error | z value | Pr(>|z|) |
|---|---|---|---|---|
| (Intercept) | 1.346 | 1.045 | 1.288 | 0.198 |
| sexM | 0.624 | 0.712 | 0.876 | 0.381 |
| flux | 0.033 | 0.058 | 0.573 | 0.567 |
| si_pp | -0.256 | 0.155 | -1.646 | 0.100 |
| si_fp | -0.185 | 0.410 | -0.452 | 0.652 |
| as.factor(sequence_pp)2 | -0.500 | 0.777 | -0.643 | 0.520 |
| sideR | -0.130 | 0.763 | -0.171 | 0.864 |

## C1) What factors affect approach latency during the familiarization phase from Day 1 to 5?

model = approach_latency ~ sex + flux + person + si + approach_latency_day0 + day + sequence

**Table S5.** Results of test of proportional hazards assumption for the Cox proportional hazards model predicting approach latency during familiarization phase from Day 1 to 5.

| Term | chisq | df | p |
|---|---|---|---|
| sex | 3.904 | 1 | 0.048 |
| as.factor(day) | 12.959 | 4 | 0.011 |
| person | 0.201 | 1 | 0.654 |
| si | 1.928 | 1 | 0.165 |
| flux | 0.681 | 1 | 0.409 |
| as.factor(sequence) | 1.120 | 1 | 0.290 |
| approach_latency_day0 | 0.147 | 1 | 0.701 |
| GLOBAL | 20.429 | 10 | 0.025 |

**Table S6.** Results of an accelerated failure time (AFT) model predicting approach latency during the familiarization phase from Day 1 to 5.

| Term | Value | Std. Error | z | p |
|---|---|---|---|---|
| (Intercept) | 1.911 | 0.094 | 20.04 | <2e-16 |
| sexM | 0.017 | 0.056 | 0.31 | 0.759 |
| personPP | -0.004 | 0.055 | -0.07 | 0.946 |
| si | -0.043 | 0.012 | -3.52 | 0.000 |
| flux | -0.000 | 0.004 | -0.10 | 0.922 |
| as.factor(sequence)2 | -0.012 | 0.055 | -0.21 | 0.832 |
| approach_latency_day0 | 0.004 | 0.002 | 1.79 | 0.073 |
| as.factor(day)2 | -0.307 | 0.091 | -3.36 | 0.001 |
| as.factor(day)3 | -0.590 | 0.087 | -6.77 | 1.3e-11 |
| as.factor(day)4 | -0.620 | 0.087 | -7.09 | 1.3e-12 |
| as.factor(day)5 | -0.760 | 0.087 | -8.74 | <2e-16 |
| Log(scale) | -1.091 | 0.040 | -27.35 | <2e-16 |

**C2) What factors affect SI during the familiarization phase from Day 1 to 5?**

model = si_new ~ sex + person + sequence + flux + si_day0 + day

**Table S7.** Results of a beta regression model predicting SI during familiarization phase from Day 1 to 5.

| Term | Estimate | Std. Error | z value | Pr(>|z|) |
| --- | --- | --- | --- | --- |
| (Intercept) | -2.878 | 0.137 | -21.027 | <2e-16 |
| sexM | -0.025 | 0.082 | -0.301 | 0.763 |
| as.factor(day)2 | 0.376 | 0.122 | 3.078 | 0.002 |
| as.factor(day)3 | 0.430 | 0.122 | 3.527 | 0.000 |
| as.factor(day)4 | 0.509 | 0.121 | 4.209 | 2.57e-05 |
| as.factor(day)5 | 0.347 | 0.123 | 2.815 | 0.005 |
| personPP | 0.092 | 0.074 | 1.277 | 0.220 |
| flux | -0.007 | 0.005 | -1.283 | 0.200 |
| as.factor(sequence)2 | -0.159 | 0.074 | -2.132 | 0.033 |
| si_day0 | 0.091 | 0.020 | 4.546 | 5.46e-06 |

**C3. What factors affect approach latency during the test phase from Day 1 to 5?**

model = approach_latency ~ sex + flux + choice + approach_latency_day0 + day + si

**Table S8.** Results of test of proportional hazards assumption for the Cox proportional hazards model predicting approach latency during test phase from Day 1 to 5.

| Term | chisq | df | p |
| --- | --- | --- | --- |
| sex | 0.491 | 1 | 0.484 |
| approach_latency_day0 | 3.675 | 1 | 0.055 |
| flux | 1.379 | 1 | 0.240 |
| as.factor(day) | 11.581 | 4 | 0.021 |
| Reward | 3.578 | 1 | 0.059 |
| SI_T | 2.403 | 1 | 0.121 |
| GLOBAL | 19.722 | 9 | 0.020 |

**Table S9.** Results of an accelerated failure time (AFT) model predicting approach latency during test phase from Day 1 to 5.

| Term | Value | Std. Error | z | p |
|---|---|---|---|---|
| (Intercept) | 2.322 | 0.154 | 15.10 | <2e-16 |
| sexM | -0.064 | 0.087 | -0.73 | 0.464 |
| flux | 0.004 | 0.006 | 0.62 | 0.534 |
| approach_latency_day0 | 0.004 | 0.003 | 1.21 | 0.225 |
| as.factor(day)2 | -0.429 | 0.142 | -3.02 | 0.002 |
| as.factor(day)3 | -0.478 | 0.138 | -3.46 | 0.000 |
| as.factor(day)4 | -0.663 | 0.140 | -4.72 | 2.4e-06 |
| as.factor(day)5 | -0.693 | 0.140 | -4.95 | 7.3e-07 |
| Reward | 0.022 | 0.088 | 0.25 | 0.803 |
| SI_T | -0.070 | 0.019 | -3.64 | 0.000 |
| Log(scale) | -0.990 | 0.056 | -17.64 | <2e-16 |

**C4. What factors affect SI during the test phase from Day 1 to 5?**

model = si_new ~ sex + flux + choice + day + si_day0

**Table S10.** Results of a beta regression model predicting SI during test phase from Day 1 to 5.

| Term | Estimate | Std. Error | z value | Pr(>|z|) |
|---|---|---|---|---|
| (Intercept) | -2.496 | 0.172 | -14.512 | <2e-16 |
| sexM | -0.183 | 0.110 | -1.659 | 0.097 |
| si_day0 | 0.094 | 0.026 | 3.667 | 0.000 |
| flux | -0.003 | 0.007 | -0.482 | 0.630 |
| as.factor(day)2 | 0.171 | 0.156 | 1.090 | 0.276 |
| as.factor(day)3 | 0.007 | 0.159 | 0.046 | 0.964 |
| as.factor(day)4 | 0.240 | 0.155 | 1.548 | 0.122 |
| as.factor(day)5 | 0.064 | 0.158 | 0.404 | 0.686 |
| Reward | -0.035 | 0.100 | -0.350 | 0.727 |

**C6. What factors affect choice from Day 1 to 5?**

model = choice ~ sex + flux + day + side + sequence + si_pp + si_fp

**Table S11.** Results of a generalized linear model with binomial distribution predicting choice of reward from Day 1 to 5.

| Term | Estimate | Std. Error | z value | Pr(>|z|) |
|---|---|---|---|---|
| (Intercept) | 1.238 | 0.604 | 2.051 | 0.040 |
| as.factor(day)2 | -0.760 | 0.454 | -1.674 | 0.094 |
| as.factor(day)3 | -0.517 | 0.458 | -1.130 | 0.258 |
| as.factor(day)4 | -0.495 | 0.458 | -1.081 | 0.280 |
| as.factor(day)5 | -0.648 | 0.449 | -1.443 | 0.149 |
| sexM | -0.099 | 0.283 | -0.349 | 0.727 |
| flux | -0.003 | 0.020 | -0.134 | 0.893 |
| si_pp | -0.151 | 0.061 | -2.484 | 0.013 |
| si_fp | 0.142 | 0.093 | 1.522 | 0.128 |
| sequence_pp | 0.008 | 0.286 | 0.026 | 0.979 |
| sideR | -0.758 | 0.285 | -2.659 | 0.008 |

**D1. What factors affect approach latency during the test phase from Day 6 to 10?**

model = approach_latency ~ sex + choice + si + flux + approach_latency_day0 + day

**Table S12.** Results of test of proportional hazards assumption for the Cox proportional hazards model predicting approach latency during test phase from Day 6 to 10.

| Term | chisq | df | p |
|---|---|---|---|
| sex | 2.867 | 1 | 0.090 |
| approach_latency_day0 | 6.123 | 1 | 0.013 |
| flux | 0.337 | 1 | 0.561 |
| as.factor(day) | 8.003 | 4 | 0.091 |
| Reward | 0.003 | 1 | 0.955 |
| SI_T | 2.516 | 1 | 0.113 |
| GLOBAL | 20.464 | 9 | 0.015 |

**Table S13.** Results of an accelerated failure time (AFT) model predicting approach latency during test phase from Day 6 to 10.

| Term | Value | Std. Error | z | p |
|---|---|---|---|---|
| (Intercept) | 1.520 | 0.160 | 9.50 | <2e-16 |
| sexM | -0.001 | 0.100 | -0.01 | 0.989 |
| approach_latency_day0 | 0.006 | 0.004 | 1.81 | 0.070 |
| flux | -0.003 | 0.007 | -0.41 | 0.685 |
| as.factor(day)7 | 0.061 | 0.138 | 0.44 | 0.151 |
| as.factor(day)8 | 0.212 | 0.148 | 1.44 | 0.151 |
| as.factor(day)9 | 0.109 | 0.142 | 0.77 | 0.444 |
| as.factor(day)10 | 0.469 | 0.148 | 3.16 | 0.002 |
| Reward | -0.052 | 0.094 | -0.55 | 0.583 |
| SI_T | -0.064 | 0.019 | -3.29 | 0.001 |
| Log(scale) | -0.912 | 0.056 | -16.23 | <2e-16 |

**D2. What factors affect SI during the test phase from Day 6 to 10?**

model = si_new ~ sex + flux + day + si_day0 + choice

**Table S14.** Results of a beta regression model predicting SI during test phase from Day 6 to 10.

| Term | Estimate | Std. Error | z value | Pr(>\|z\|) |
|---|---|---|---|---|
| (Intercept) | -2.124 | 0.148 | -14.307 | <2e-16 |
| sexM | -0.319 | 0.101 | -3.154 | 0.002 |
| flux | -0.014 | 0.006 | -2.201 | 0.028 |
| as.factor(day)7 | 0.142 | 0.146 | 0.974 | 0.330 |
| as.factor(day)8 | 0.198 | 0.145 | 1.364 | 0.172 |
| as.factor(day)9 | 0.052 | 0.148 | 0.354 | 0.724 |
| as.factor(day)10 | 0.027 | 0.148 | 0.181 | 0.856 |
| Reward | -0.179 | 0.094 | -1.894 | 0.058 |
| si_day0 | 0.106 | 0.025 | 4.239 | 2.24e-05 |

**D4. What factors affect choice from Day 6 to 10?**

model = choice ~ sex + flux + day + side

**Table S15.** Results of a generalized linear model with binomial distribution predicting choice of reward from Day 6 to 10.

| Term | Estimate | Std. Error | z value | Pr(>\|z\|) |
|---|---|---|---|---|
| (Intercept) | 1.043 | 0.849 | 1.229 | 0.219 |
| sexM | -0.428 | 0.927 | -0.461 | 0.645 |
| flux | 0.010 | 0.095 | 0.102 | 0.919 |
| as.factor(day)7 | -0.970 | 0.989 | -0.981 | 0.326 |
| as.factor(day)8 | -0.516 | 0.984 | -0.524 | 0.600 |
| as.factor(day)9 | 0.609 | 1.080 | 0.564 | 0.573 |
| as.factor(day)10 | -0.396 | 0.993 | -0.398 | 0.690 |
| sideR | -0.353 | 0.690 | -0.512 | 0.609 |